\documentclass[aps,prb,showpacs,showkeys,twocolumn,longbibliography]{revtex4-1}  

\usepackage{amsmath,amssymb}
\usepackage{graphicx}
\begin{document}
\newcommand{\dd}{\mathrm{d}}
\newcommand{\ii}{\mathrm{i}}
\newcommand{\ee}{\mathrm{e}}
\newcommand{\Sp}{\mathrm{Sp}}
\newcommand{\heav}{{\text{\usefont{U}{psy}{m}{n}\selectfont\symbol{113}}}}
\newcommand{\dirac}{{\text{\usefont{U}{psy}{m}{n}\selectfont\symbol{100}}}}
\newcommand{\kronek}{{\text{\usefont{U}{psy}{m}{n}\selectfont\symbol{100}}}}
\newcommand{\lindhard}{\mathrm{L}}
\newcommand{\jell}{\mathrm{jell}}
\newcommand{\unif}{\mathrm{unif}}
\newcommand{\bulk}{\mathrm{bulk}}
\newcommand{\surf}{\mathrm{surf}}
\newcommand{\slab}{\mathrm{slab}}






\title{Effect of Coulomb interaction on chemical potential of metal film}


\author{P. P. Kostrobij, B. M. Markovych}
\email[]{bogdan\_markovych@yahoo.com}
\affiliation{Lviv Polytechnic National University, 12 Bandera Str., 79013 Lviv, Ukraine}


\date{\today}

\begin{abstract}
 The chemical potential of a metal film within the jellium model
 with taking into account the Coulomb interaction between electrons is calculated.
 The surface potential is modeled as the infinite rectangular potential well.
 The behavior of the chemical potential as a function of the film thickness is studied,
 the quantum size effect for this quantity is discovered.
 It is shown
 that taking into account the Coulomb interaction leads to
 a significant decrease of the chemical potential and
 to an enhancement of the quantum size effect.
\end{abstract}


\maketitle


 \section{Introduction}
 The rapid development of nanotechnology involving processes of metal deposition on various substrates
 requires a theoretical analysis and understanding of electronic effects in nanoclusters and nanofilms.
 If the size of nanostructure is comparable with the corresponding Fermi wavelength of electrons in the nanostructure,
 various physical properties may strongly depend on the size of this nanostructure.
 This phenomenon is called the quantum size effect~\cite{Cohen199042,Heer1993611,Brack1993677,Chiang2000181,Milun200299,Tringides200750}
 and is typical for many physical quantities of metal nanofilms,
 such as thermodynamic stability, electrical resistivity, work function, surface energy, etc~\cite{Yong2009155404}.
 Due to possible differences in properties of metal nanostructures from properties of the bulk metal,
 the research of such properties has considerable theoretical and experimental interests.

 The first theoretical calculations of the chemical potential of the metal film
 within the jellium model without taking into account the Coulomb interaction between electrons are presented
 in Refs.~[\onlinecite{Thompson19636,Smith1965210,Paskin1965A1965}].
 However, as it is shown in Refs.~[\onlinecite{Stratton1965556,Schulte1977149}],
 these calculations do not take into account the condition of electroneutrality and
 therefore the calculated values of the chemical potential are not correct.
 A few years ago Ref.~[\onlinecite{Dymnikov2011901}] appeared,
 the author of which claimed that he was the first who found the dependence of the Fermi energy on the film thickness
 within the same model of a metal film.
 However, in fact, it is a repetition of some results of Ref.~[\onlinecite{Thompson19636}] without reference to them.

 For the first time the quantum size effect on the chemical potential in the metal film within
 the jellium model was studied by Shulte~\cite{Schulte1976427} using the density functional theory within the local
 density approximation.
 Later,
 he compared it with the magnitudes of the chemical potential of the film within the jellium model
 without the Coulomb interaction between electrons, which are also obtained by him~\cite{Schulte1977149},
 and found good agreement.
 However,
 in his calculations \cite{Schulte1977149},
 Schulte used for the distance between the side of the film and the potential wall the result of Ref.~[\onlinecite {Bardeen1936653}],
 which is true for the semi-infinite jellium
 (see, for example, Refs.~[\onlinecite{Bardeen1936653,Huntington19511035,Kostrobij2015075441}]).
 This result is also used in much later Refs.~[\onlinecite{Pitarke2001045116,Yong2009155404}],
 in which there is also a comparison between calculations by the density functional theory and
 the calculations within the jellium model without the Coulomb interaction,
 and good agreement is found for the specific films~\cite{Yong2009155404}.
 The correct expression for this distance is obtained in Ref.~[\onlinecite{Himbergen19782674}],
 where analytical calculations for various models of potential barrier within the jellium model
 without the Coulomb interactions between electrons are conducted.
 In Ref.~[\onlinecite{Biao2008035410}],
 the chemical potential and the stability of metal thin film within the jellium model
 without the Coulomb interaction between electrons are studied,
 however,
 as in Refs.~[\onlinecite{Thompson19636,Smith1965210,Paskin1965A1965,Dymnikov2011901}],
 the condition of electroneutrality is not taken into account.

 In the present work,
 the metal film within the jellium model taking into account the Coulomb interactions between electrons is studied.
 The surface potential is modeled by the infinite rectangular potential well.
 In the limit of low temperatures,
 calculations of the chemical potential and
 the distance between the side of the film and the potential wall of infinite height are performed
 for different values of the Wigner-Seitz radius ($r_\mathrm{s}$).
 The chemical potential is found as a solution of the nonlinear equation,
 which is obtained in Ref.~[\onlinecite{Kostrobij2015075441}]
 by using the method of functional integration.
 The dependences of the calculated quantities on the film thickness are studied,
 it is shown that taking into account the Coulomb interaction between electrons leads
 to a significant decrease in the chemical potential and increase in the distance between a side of the film and the infinite potential wall,
 and to an increase of the amplitudes of its oscillations,
 i.e. to an enhancement of the quantum size effect.
 It is shown that if the film thickness increases,
 the chemical potential of the film tends to
 the bulk chemical potential,
 i.e. to the chemical potential of unbounded metal within the jellium model,
 and the distance tends to magnitude,
 which  is obtained in Ref.~[\onlinecite{Kostrobij2015075441}] for the semi-infinite jellium.

 \section{Model}
\begin{figure}[hbtp]
  \centering
  \includegraphics[width=4.5cm]{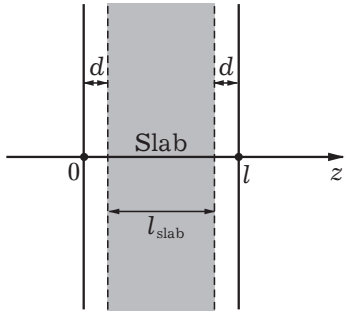}\\
  \caption{Scheme of film.
           Two vertical dashed lines represent the geometrical surfaces of the film,
           two vertical solid lines represent the surface potential boundaries.}\label{Slab}
\end{figure}

 We consider a metal slab
 placed in such way
 that its two parallel infinite sides are parallel to the $xOy$ plane.
 Thickness of the slab is denoted by $l_\slab$ and lies along the $z$ axis.
 One side of the slab is specified by the equation ${z=d}$,
 and the second one is described by the equation ${z=l_\slab+d}$ as shown in Fig.~\ref{Slab}.

 The slab is considered within the jellium model,
 i.e. an ionic subsystem is replaced by positive charge
 with the distribution
 \begin{align*}
   \varrho_\jell(\mathbf{r}_{||},z)
   &\equiv
   \varrho_\jell(z)
   =\varrho_0\,\heav(z-d)\,\heav(l_\slab+d-z)\\
   &=\left\{
      \begin{array}{ll}
        \varrho_0, & z\in[d,l_\slab+d], \\
        0, & z\not\in[d,l_\slab+d],
      \end{array}
    \right.
 \end{align*}
 where $\heav(x)$ is the Heaviside step function,
 ${\textbf{r}_{||}=(x,y)}$,
 ${x,y\in(-\sqrt{S}/2,+\sqrt{S}/2)}$,
 ${z\in(-\infty,+\infty)}$,
 $S$ is area of the side of the slab (${S\to\infty}$).
 The condition of electroneutrality is satisfied,
 \begin{equation}\label{electroNeutr}
  \lim_{S\to\infty}
  \int_S \dd{\bf r}_{||}
  \int_{-\infty}^{+\infty}  \dd z \,
  \varrho_\jell({\bf r}_{||},z) = e N, \; e>0
 \end{equation}
 moreover, in the thermodynamic limit, we have
 \begin{equation*}
    \lim_{N,S\to\infty}\frac{eN}{Sl_\slab}=\varrho_0,
 \end{equation*}
 where $N$ is the number of electrons,
 which are situated in the field of the positive charge.
 The parameter $d$ is determined by the condition of electroneutrality.

 As a consequence of the symmetry of the model,
 the motion of the electron in a plane parallel to the $xOy$ plane is free,
 and the one along the $z$ axis is determined by the surface potential $V_\surf(z)$.
 This potential is modeled by the infinite rectangular potential well,
 namely,
 \begin{equation}\label{Vsurf}
   V_\surf(z)
   =\left\{
      \begin{array}{cc}
        \infty, & z\leqslant0,\;z\geqslant l, \\
        0, & 0< z< l.
      \end{array}
    \right.
 \end{equation}
 This model potential allows an analytical solving of the Schr\"odinger stationary equation,
 \begin{equation*}
   \left[
    -\frac{\hbar^2}{2m}\Delta+V_\surf(z)
   \right]\Psi_a(\mathbf{r})
   =E_a\Psi_a(\mathbf{r}),\quad
   \mathbf{r}=(\mathbf{r}_{||},z)
 \end{equation*}
 with the Dirichlet boundary conditions,
 \[
  \Psi_a(\mathbf{r})\big|_{z=0}=
  \Psi_a(\mathbf{r})\big|_{z=l}=0,
 \]
 where $m$ is the electron mass,
 ${a=(\mathbf{k}_{||},\alpha)}$,
 $\mathbf{k}_{||}$ is  the two-dimensional wave vector of the electron in the plane parallel to the $xOy$ plane,
 ${\alpha=\pi n/l}$,
 ${n=1,2,\ldots}$.
 The wave functions and the corresponding energy levels for the potential model~\eqref{Vsurf} are
 \[
  \Psi_a(\mathbf{r})
  =
  \frac1{\sqrt S}\ee^{\ii \mathbf{k}_{||} \mathbf{r}_{||}}
  \varphi_\alpha(z),
 \]
\begin{align}
    \varphi_\alpha(z)&=
  \sqrt{\frac2l}\sin(\alpha z)\,
  \heav(z)\,
  \heav(l-z)\nonumber\\
  &=\left\{
    \begin{array}{cc}
      \sqrt{\frac2l}\sin(\alpha z), & 0<z<l,\\
      0, & z\leqslant0,\;z\geqslant l.
    \end{array}
  \right.\label{wavef}
\end{align}
 \[
  E_a=\frac{\hbar^2(k_{||}^2+\alpha^2)}{2m}.
 \]

 As we see from Fig.~\ref{Slab},
 there is the relation between the parameter $l$ of the model potential
 and the thickness $l_\slab$ of the slab
 \begin{equation}\label{relatdl}
  l=l_\slab+2d,
 \end{equation}
 where the parameter $d$ is found in Ref.~[\onlinecite{Himbergen19782674}],
 \begin{equation}\label{d}
   d=\frac{3\pi}{8\mathcal{K}_\mathrm{F}}+\frac{\pi^2}{8\mathcal{K}_\mathrm{F}^2l},
 \end{equation}
 ${\mathcal{K}_\mathrm{F}=\sqrt{2m\mu}/\hbar}$ is the magnitude of the Fermi wave vector,
 $\mu$ is the chemical potential.

 From Eqs.~\eqref{d} and \eqref{relatdl},
 we find the parameter $l$ of the infinite rectangular potential well
 as function of $\mathcal{K}_\mathrm{F}$,
 \[
  l(\mathcal{K}_\mathrm{F})=\frac{l_\slab}{2}+\frac{3\pi}{8\mathcal{K}_\mathrm{F}}+
     \frac{\sqrt{16\mathcal{K}_\mathrm{F}^2l_\slab^2+24\pi\mathcal{K}_\mathrm{F}l_\slab+25\pi^2}}{8\mathcal{K}_\mathrm{F}}.
 \]

 It should be noted that
 if the parameter $l$ approaches infinity,
 the parameter~$d$ approaches the well-known magnitude ${d_{\infty}=3\pi/(8\mathcal{K}_\mathrm{F})}$,
 which is the distance between locations of the edge of the positive charge and the infinite potential wall
 within the jellium model
 (see, for example, Ref.~[\onlinecite{Bardeen1936653,Huntington19511035,Kostrobij2015075441}]).

 \section{Equation for the chemical potential}

 In Ref.~[\onlinecite{Kostrobij2015075441}],
 using the method of functional integration,
 the general expression for the average number operator of electrons within the semi-infinite jellium model is obtained.
 In the case of the slab,
 this expression has the form
  \begin{align}\label{N1}
    \langle N\rangle&=\langle N\rangle_0-\frac1{2S}\sum\limits_{\mathbf{q}\neq0}
                     \sum\limits_{\mathbf{k}_{||},\alpha}
                     \frac{\partial n_\alpha(\mathbf{k}_{||})}{\partial\mu}\nonumber\\
      &\qquad \times\int_{0}^{l}\!\dd z\,|\varphi_\alpha(z)|^2
      \big(g(\mathbf{q},z,z)-\nu(\mathbf{q},0)\big)\nonumber\\
    &\quad+\frac1{2S}\sum\limits_{\mathbf{q}\neq0}
                     \sum\limits_{\mathbf{k}_{||},\alpha_1,\alpha_2}
      \frac{\partial \big(n_{\alpha_1}(\mathbf{k}_{||}) \,
      n_{\alpha_2}(\mathbf{k}_{||}-\mathbf{q})\big)}{\partial\mu}\nonumber\\
      &\qquad \times
      \int_{0}^{l}\!\dd z_1
      \int_{0}^{l}\!\dd z_2\,
      \varphi^*_{\alpha_1}\!(z_1) \varphi^{\vphantom{*}}_{\alpha_2}\!(z_1)\nonumber\\
      &\quad\qquad \times
      \varphi^*_{\alpha_2}\!(z_2) \varphi^{\vphantom{*}}_{\alpha_1}\!(z_2)
      \,g(\mathbf{q},z_1,z_2),
 \end{align}
 where
 \begin{equation}\label{N0}
   \langle N\rangle_0=\sum\limits_{\mathbf{k}_{||},\alpha}n_\alpha(\mathbf{k}_{||})
 \end{equation}
 is the average number operator of noninteracting electrons
 (i.e. without the Coulomb interaction between electrons),
 ${
  n_\alpha(\mathbf{k}_{||})=\frac{1}{\ee^{\beta(E_\alpha(\mathbf{k}_{||})-\mu)}+1}}
 $
 is the Fermi-Dirac distribution,
 $\beta$ is the inverse thermodynamic temperature,
 ${\mathbf{q}=(q_x,q_y)}$ is the two-dimensional vector with components
 ${q_{x,y}=2\pi n_{x,y}/\sqrt{S}}$,
 ${n_{x,y}=0,\pm1,\pm2,\ldots}$,
 ${\nu(\mathbf{q},0)=2\pi e^2/q}$,
 $g(\mathbf{q},z_1,z_2)$ is the effective interelectron interaction in $(\mathbf{q},z)$ representation,
 analytical expression for which is obtained in Ref.~[\onlinecite{Kostrobij201651}]
 within the same level of approximations used in Refs.~[\onlinecite{Kostrobij2015075441,Kostrobij2016155401}].

 In the limit of low temperatures,
 we obtain
 $
  n_\alpha(\mathbf{k}_{||})=\heav\big(\mathcal{K}_\mathrm{F}^2-k^2_{||}-\alpha^2\big).
 $
 By using transition from the summation over $\mathbf{k}_{||}$ to the integration according to the rule~\cite{Kostrobij2015075441,Kostrobij2016155401},
 \[
  \sum_{\mathbf{k}_{||}}\ldots
  =\frac{2S}{(2\pi)^2}
   \int \!\dd\mathbf{k}_{||}\ldots,
 \]
 where two possible orientations of the electron spin are taken into account,
 a summation over the two-dimensional vector $\mathbf{k}_{||}$ in Eqs.~\eqref{N1} and \eqref{N0} can be performed analytically.
 As a result,
 we find that
\begin{align*}
     &\sum_{\mathbf{k}_{||}}n_\alpha(\mathbf{k}_{||})
   =
   \frac{S}{2\pi}\big(\mathcal{K}^2_\mathrm{F}-\alpha^2\big)\,
   \heav\big(\mathcal{K}^2_\mathrm{F}-\alpha^2\big), \\
  &\sum_{\mathbf{k}_{||}}\frac{\partial n_\alpha(\mathbf{k}_{||})}{\partial \mu}
   =
   \frac{S}{2\pi}\frac{2m}{\hbar^2}\,
   \heav\big(\mathcal{K}^2_\mathrm{F}-\alpha^2\big).
\end{align*}
 In Ref.~[\onlinecite{Kostrobij2015075441}],
 it is shown that
 \begin{equation*}
   \sum\limits_{\mathbf{k}_{||}}
    \frac{\partial\big( n_{\alpha_1}(\mathbf{k}_{||})\,n_{\alpha_2}(\mathbf{k}_{||}-\mathbf{q})\big)}{\partial\mu}
    =\frac{2S}{(2\pi)^2}\frac{4m}{\hbar^2}I(q,\alpha_1,\alpha_2),
 \end{equation*}
 where the expression for the function $I(q,\alpha_1,\alpha_2)$ is given in Ref.~[\onlinecite{Kostrobij2015075441}].

 A summation over the quantum numbers $\alpha$ can be represented as
 $
  \sum_\alpha\ldots
  =
  \sum_{n=1}^{n_{\max}}\ldots,
 $
 where $n_{\max}$ is the integer part of ${[l\mathcal{K}_\mathrm{F}/\pi]}$.

 As shown in Ref.~[\onlinecite{Thompson19636}],
 the calculation of the average number operator of noninteracting electrons~\eqref{N0}
 can be performed analytically,
 \begin{equation*}\label{N02}
   \langle N\rangle_0
   =
   \frac{S}{2\pi}n_{\max}
   \left(
    \mathcal{K}_\mathrm{F}^2
    -
    \frac{\pi^2}{6l^2}(n_{\max}+1)(2n_{\max}+1)
   \right).
 \end{equation*}

 From the condition of electroneutrality~\eqref{electroNeutr}
 it follows that ${e\langle N\rangle=\varrho_0Sl_\slab}$,
 i.e. ${\varrho_0=e\langle N\rangle/(Sl_\slab)}$.
 If we assume that the concentration of the positive charge is equal to
 the electron concentration of unbounded metal,
 i.e. ${\varrho_0/e=3/(4\pi r^3_\mathrm{s})}$,
 we have
 \begin{equation}\label{concen}
  \frac{3}{4\pi r^3_\mathrm{s}}=\frac{\langle N\rangle}{Sl_\slab},
 \end{equation}
 where $r_\mathrm{s}$ is the Wigner-Seitz radius.
 By multiplying Eq.~\eqref{N1} by ${2\pi/(Sl_\slab)}$ and using Eq.~\eqref{concen},
 we obtain the nonlinear algebraic equation for the magnitude of the Fermi wave vector~$\mathcal{K}_\mathrm{F}$,
 which is connected with the chemical potential~$\mu$, ${\mu=\hbar^2\mathcal{K}^2_\mathrm{F}/(2m)}$,
 (integrals of the effective interelectron interaction and the wave functions~\eqref{wavef} are calculated in Appendix~\ref{intG}),
 \begin{widetext}
 \begin{align}\label{EqForChemPotEnd}
   \frac{3}{2r_\mathrm{s}^3} & = \frac{n_{\max}}{l_\slab}\left(
    \mathcal{K}_\mathrm{F}^2
    -
    \frac{\pi^2}{6l^2}(n_{\max}+1)(2n_{\max}+1)
   \right)\nonumber\\[-1mm]
    &\quad-\frac{a_\mathrm{B}^2}{l_\slab}\sum_{n=1}^{n_{\max}}
      \int\limits_0^\infty\!\dd q
       \Bigg[\frac qQ
       \frac{1}{1-\left(\frac{Q-q}{Q+q}\right)^2\ee^{-2Ql}}
       \Bigg(1
   +\left(\frac{Q-q}{Q+q}\right)^2\ee^{-2Ql}+\frac{4\alpha^2}{Ql}\frac{Q-q}{Q+q}\frac{1-\ee^{-2Ql}}{4\alpha^2+Q^2}\Bigg)-1\Bigg]\nonumber\\[-1mm]
   &\quad+\frac{8}{\pi}\frac{a_\mathrm{B}^4}{l_\slab l^2}\sum_{n_1=1}^{n_{\max}} \sum_{n_2=1}^{n_{\max}}
     \int\limits_0^\infty\!\dd q\,\frac qQ
     \frac{1}{1-\left(\frac{Q-q}{Q+q}\right)^2\ee^{-2Ql}}\nonumber\\[-2mm]
   &\qquad\times   \Bigg[ I_1(Q,\alpha_1,\alpha_2)  +\left(\frac{Q-q}{Q+q}\right)^2\ee^{-2Ql}I_1(-Q,\alpha_1,\alpha_2)+
   \frac{Q-q}{Q+q}\left(I_2^2(Q,\alpha_1,\alpha_2)+\ee^{-2Ql}I_2^2(-Q,\alpha_1,\alpha_2)\right) \Bigg],
 \end{align}
 where $a_\mathrm{B}$ is the Bohr radius,
 expressions for $Q$, functions $I_1$, and $I_2$ are given in Appendix~\ref{intG}
 (see Eqs.~\eqref{exprQ}, \eqref{I1}, and \eqref{I2}, respectively).
 \end{widetext}

 It should be noted that
 in the case of noninteracting electrons,
 the nonlinear equation~\eqref{EqForChemPotEnd}
 is significantly simplified,
 \begin{equation}\label{EqForChemPot0End}
   \frac{3}{2r_\mathrm{s}^3}  = \frac{n_{\max}}{l_\slab}\left(
    (\mathcal{K}_\mathrm{F}^0)^2
    -
    \frac{\pi^2}{6l_0^2}(n_{\max}+1)(2n_{\max}+1)
   \right),
  \end{equation}
 where $\mathcal{K}_\mathrm{F}^0$ is the magnitude of the Fermi wave vector of noninteracting electrons,
 ${l_0=l(\mathcal{K}_\mathrm{F}^0)}$.
 If we solve this equation,
 we determine the chemical potential ${\mu^0=\hbar^2(\mathcal{K}_\mathrm{F}^0)^2/(2m)}$ of noninteracting electrons.

 \section{Results of the numerical calculations and discussion}
\begin{figure}[htbp]
  \centering
  \includegraphics[width=7.8cm]{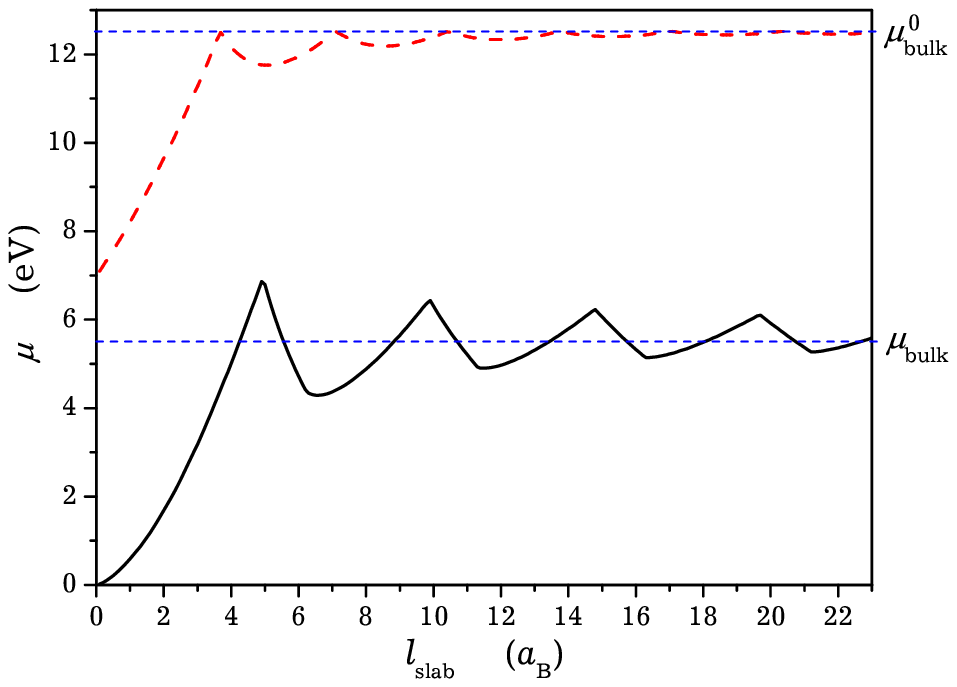}\\[2mm]
  \includegraphics[width=7.8cm]{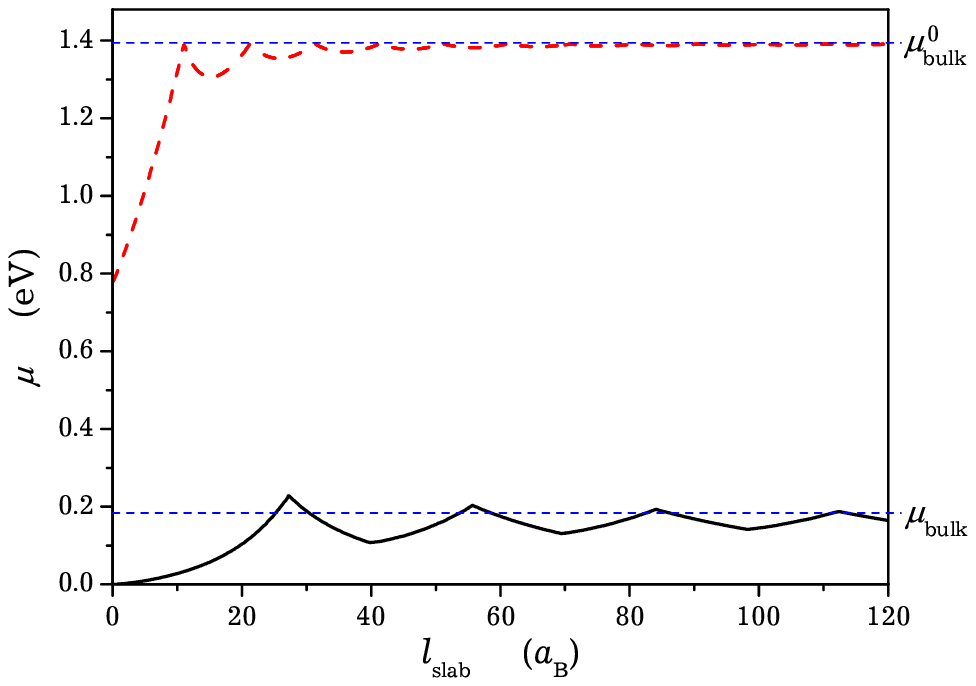}\\[-2mm]
  \caption{The chemical potential as a function of the film thickness at ${r_\mathrm{s}=2a_\mathrm{B}}$ (top) and ${r_\mathrm{s}=6a_\mathrm{B}}$ (buttom).}\label{mu}
\end{figure}

 In Fig.~\ref{mu},
 the chemical potential as a function of the film thickness is presented
 for the following values of the Wigner-Seitz radius:
 ${r_\mathrm{s}=2a_\mathrm{B}}$ and ${r_\mathrm{s}=6a_\mathrm{B}}$.
 The solid curve represents the chemical potential with taking into account the Coulomb interaction between electrons,
 i.e. that is found from the nonlinear algebraic equation~\eqref{EqForChemPotEnd},
 the dashed curve represents one without this interaction,
 i.e. that is found from the nonlinear algebraic equation~\eqref{EqForChemPot0End}.
 In addition,
 the short-dashed horizontal curves show the bulk chemical potential~\cite{Kostrobij2015075441}
 with taking into account the Coulomb interaction between electrons $\mu_\bulk$ and one without this interaction $\mu_\bulk^0$,
 respectively.

 We see that the dependence of the chemical potential on the film thickness is non monotonic,
 there are alternating peaks,
 i.e. we observe the quantum size effect for the chemical potential of the metal film.
 This is a consequence of quantization of the electron energy levels,
 because the motion of electrons in the direction perpendicular to the film is limited.
 If the film thickness  increases,
 the quantum size effect vanishes,
 and the chemical potential tends to the bulk chemical potential.
 Taking into account the Coulomb interaction between electrons leads to a significant decrease in the chemical potential,
 what is known (see, for example, Ref.~[\onlinecite{Kostrobij2015075441}]),
 and also to an enhancement of the quantum size effect:
 peaks become higher and valleys become deeper,
 and the period of alternating peaks and valleys increases.
 As noted by Schulte~\cite{Schulte1976427},
 the distance between adjacent peaks of the chemical potential
 without the Coulomb interaction between electrons is about ${\lambda_\mathrm{F}^0/2}$,
 where ${\lambda_\mathrm{F}^0 =2\pi/\mathcal{K}_\mathrm{F}^0}$ is the Fermi wavelength of noninteracting electrons.
 It turns out that
 if the Coulomb interaction between electrons is taken into account,
 the distance between adjacent peaks of the chemical potential is also ${\lambda_\mathrm{F}/2}$,
 but here ${\lambda_\mathrm{F}=2\pi/\mathcal{K}_\mathrm{F}}$ is the Fermi wavelength of interacting electrons.

\begin{figure}[htbp]
  \centering
  \includegraphics[width=7.8cm]{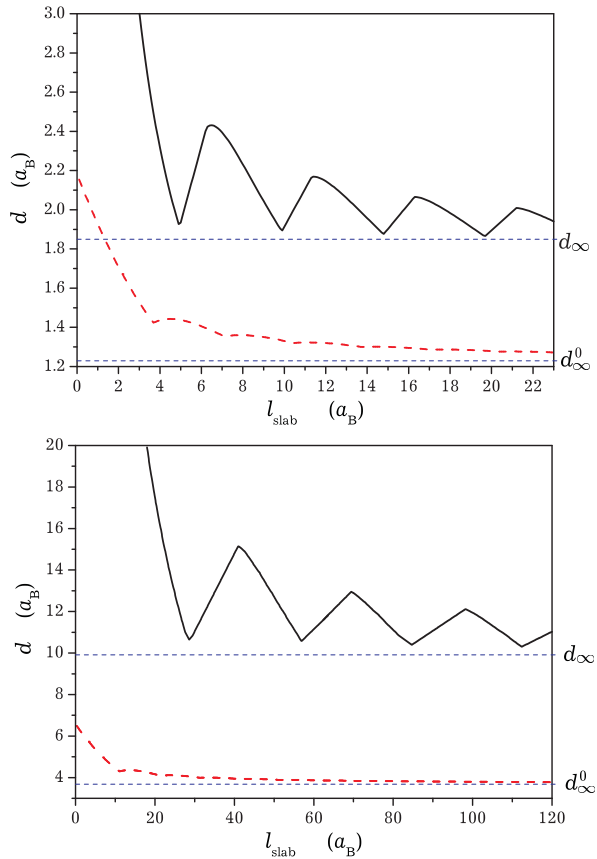}\\[-2mm]
  \caption{The parameter $d$ as a function of the film thickness at ${r_\mathrm{s}=2a_\mathrm{B}}$ (top) and ${r_\mathrm{s}=6a_\mathrm{B}}$ (buttom).}\label{parameterd}
\end{figure}

 As noted above,
 the authors of Refs.~[\onlinecite{Thompson19636,Smith1965210,Dymnikov2011901,Atkinson20081099,Biao2008035410}] do not consider
 the parameter $d$ in calculating the chemical potential of the metal film without Coulomb interaction between electrons,
 i.e. they believed that positions of the potential wall and the edge of the positive charge coincide.
 As a result,
 the chemical potential calculated by them  is too large,
 moreover if the film thickness increases,
 the chemical potential does not tend to the bulk chemical potential.
 For the first time,
 the chemical potential of the metal film without the Coulomb interaction between electrons,
 but with taking into account the parameter $d$ was calculated by Schulte~\cite{Schulte1977149}.
 However,
 he used the magnitude ${d_\infty^0=3\pi/(8\mathcal{K}_\mathrm {F}^0)}$ for the parameter $d$ instead of Eq.~\eqref{d},
 i.e. he used the magnitude for the semi-infinite jellium
 (see., for example, Refs.~[\onlinecite{Bardeen1936653,Huntington19511035,Kostrobij2015075441}]).
 This led to a very good agreement of the chemical potential obtained without the Coulomb interaction between electrons
 and the chemical potential calculated by Schulte~\cite{Schulte1976427} before
 with using the density functional theory within the local density approximation.
 Such agreement is strange,
 because the presented calculations and the results of our previous work~\cite{Kostrobij2015075441}
 show that the Coulomb interactions between electrons leads to a significant decrease in the chemical potential.
 As can be seen from the analysis of results for the chemical potential,
 the correct calculation of the parameter $d$ is very important.
 In Fig.~\ref{parameterd},
 the parameter~$d$ as a function of the film thickness is presented
 for the following values of the Wigner-Seitz radius:
 ${r_\mathrm{s}=2a_\mathrm{B}}$ and ${r_\mathrm{s}=6a_\mathrm{B}}$.
 The solid curve represents the parameter $d$ with taking into account the Coulomb interaction between electrons,
 the dashed curve represents one without this interaction.
 In addition,
 the short-dashed horizontal lines show the parameter $d$ of the semi-infinite jellium model,
 i.e. at ${l_\slab\to\infty}$,
 with taking into account the Coulomb interaction between electrons $d_\infty$ and one without this interaction $d_\infty^0$,
 respectively.

 We see that the dependence of the parameter $d$ on the film thickness is non monotonic,
 there are alternating peaks,
 i.e. we observe also the quantum size effect for the parameter $d$.
 If the film thickness increases,
 the quantum size effect vanishes.
 Taking into account the the Coulomb interaction between electrons leads to a significant increase in the parameter $d$,
 what is known from Ref.~[\onlinecite {Kostrobij2015075441}],
 and also to an enhancement of the quantum size effect:
 peaks become higher and valleys become deeper,
 and the period of alternating peaks and valleys increases.
 As it is for the chemical potential,
 the period of alternating is about ${\lambda_\mathrm{F}^0/2}$ in the absence of the Coulomb interaction between electrons and
 ${\lambda_\mathrm{F}/2}$ in the presence of it.

 \section{CONCLUSIONS}
 In the limit of low temperatures, by solving the nonlinear algebraic equations,
 the chemical potential of the metal film within the jellium model
 with taking into account the Coulomb interaction between the electrons is calculated.
 It is shown that
 the dependence of the chemical potential on the film thickness is non monotonic,
 i.e. there is the quantum size effect for the chemical potential of the metal film,
 the distance between neighboring maxima is about half of the Fermi wavelength.
 If the film thickness increases,
 the quantum size effect disappears and
 the chemical potential tends to the bulk chemical potential.
 In addition,
 taking into account the Coulomb interaction between electrons enhances the quantum size effect and
 leads to a significant decrease in the chemical potential.

 In the same approximations,
 the parameter $d$,
 which is the distance from the side of the film to the potential wall,
 is calculated as the function of the film thickness.
 It is shown that
 this dependence on the film thickness is also non monotonic,
 there is the quantum size effect of the parameter $d$.
 If the film thickness increases, this quantum size effect disappears also.
 Taking into account the Coulomb interaction between electrons leads
 to a significant increase in the parameter $d$, and to an enhancement of the quantum size effect.
 The distance between neighboring maxima is also about half of the Fermi wavelength.

 \appendix
  \renewcommand{\theequation}{\Alph{section}\arabic{equation}}

\begin{widetext}
 \section{CALCULATION OF INTEGRALS WITH EFFECTIVE INTERELECTRON INTERACTION}\label{intG}
 An expression for the effective interelectron interaction $g(\mathbf{q},z_1,z_2)$ in $(\mathbf{q},z)$ representation
 is obtained in Ref.~[\onlinecite{Kostrobij201651}].
 In the domain ${0\leqslant z_1,z_2\leqslant l}$,
 it has the form
 \begin{equation}\label{geff}
     g(q|z_1,z_2)=\frac{2\pi e^2}{Q}\frac{1}{1-\left(\frac{Q-q}{Q+q}\right)^2\ee^{-2Ql}}
   \Bigg[\ee^{-Q|z_1-z_2|}
   +\left(\frac{Q-q}{Q+q}\right)^2\ee^{-Q(2l-|z_1-z_2|)} +\frac{Q-q}{Q+q}\Big(\ee^{-Q(z_1+z_2)}+\ee^{-Q(2l-z_1-z_2)}\Big)\Bigg],
 \end{equation}
 where
 \begin{equation}\label{exprQ}
  Q=\sqrt{q^2+\varkappa^2},
 \end{equation}
 ${\varkappa^2(q)
  =
  \frac{4}{la_\mathrm{B}}\sum\limits_\alpha}
  \Bigg[1-\sqrt{1-4\frac{\mathcal{K}^2_\mathrm{F}-\alpha^2}{q^2}}\,
  \heav\left(1-4\frac{\mathcal{K}^2_\mathrm{F}-\alpha^2}{q^2}\right)\Bigg]\,
  \heav(\mathcal{K}_\mathrm{F}-\alpha)$,
 $a_\mathrm{B}$ is the Bohr radius.

 The integrals of products of the wave functions~\eqref{wavef} and the effective interelectron interaction~\eqref{geff} in Eq.~\eqref{N1}
 can be analytically calculated,
 and we obtain
 \[
   \int_{0}^{l}\!\dd z\,|\varphi_\alpha(z)|^2g(\mathbf{q},z,z)  = \frac{2\pi e^2}{Q}\frac{1}{1-\left(\frac{Q-q}{Q+q}\right)^2\ee^{-2Ql}}
   \Bigg[1
   +\left(\frac{Q-q}{Q+q}\right)^2\ee^{-2Ql}+\frac{4\alpha^2}{Ql}\frac{Q-q}{Q+q}\frac{1-\ee^{-2Ql}}{4\alpha^2+Q^2}\Bigg],\vspace{-6mm}
 \]
 \begin{multline*}
    \int_{0}^{l}\dd z_1\!\int_{0}^{l}\dd z_2\,
      \varphi^*_{\alpha_1}\!(z_1) \varphi^{\vphantom{*}}_{\alpha_2}\!(z_1)
      \varphi^*_{\alpha_2}\!(z_2) \varphi^{\vphantom{*}}_{\alpha_1}\!(z_2)
      \,g(\mathbf{q},z_1,z_2)
      =\frac{4}{l^2}\frac{2\pi e^2}{Q}\frac{1}{1-\left(\frac{Q-q}{Q+q}\right)^2\ee^{-2Ql}}\\[-2mm]
      \times\Bigg[
     I_1(Q,\alpha_1,\alpha_2)
     +\left(\frac{Q-q}{Q+q}\right)^2\ee^{-2Ql}I_1(-Q,\alpha_1,\alpha_2)
     +\frac{Q-q}{Q+q}\left(I_2^2(Q,\alpha_1,\alpha_2)+\ee^{-2Ql}I_2^2(-Q,\alpha_1,\alpha_2)\right) \Bigg],
 \end{multline*}
 where
 \begin{align}
   I_1(Q,\alpha_1,\alpha_2)&=\frac{8Q^2\alpha_1^2\alpha_2^2\left(\ee^{-Ql}\cos(\alpha_1l)\cos(\alpha_2l)-1\right)}
                                {\left[\left(Q^2+\alpha_1^2+\alpha_2^2\right)^2-4\alpha_1^2\alpha_2^2\right]^2}+
                                \frac{l\alpha_1^2\dirac_{\alpha_1,\alpha_2}}{Q(8\alpha_1^2+2Q^2)},\label{I1}\\
  I_2(Q,\alpha_1,\alpha_2)&=\frac{2Q\alpha_1\alpha_2\left(1-\ee^{-Ql}\cos(\alpha_1l)\cos(\alpha_2l)\right)}
                                {\left(Q^2+\alpha_1^2+\alpha_2^2\right)^2-4\alpha_1^2\alpha_2^2} .\label{I2}
 \end{align}
\end{widetext}



\begin{thebibliography}{10}

\bibitem{Atkinson20081099}
W.~A. Atkinson and A.~J. Slavin.
\newblock A free-electron calculation for quantum size effects in the
  properties of metallic islands on surfaces.
\newblock {\em American Journal of Physics}, 76(12):1099--1101, 2008.

\bibitem{Bardeen1936653}
J.~Bardeen.
\newblock {Theory of the Work Function. II. The Surface Double Layer}.
\newblock {\em Phys. Rev.}, 49(9):653--663, 1936.

\bibitem{Brack1993677}
M.~Brack.
\newblock The physics of simple metal clusters: self-consistent jellium model
  and semiclassical approaches.
\newblock {\em Rev. Mod. Phys.}, 65(3):677--732, 1993.

\bibitem{Chiang2000181}
T.-C. Chiang.
\newblock Photoemission studies of quantum well states in thin films.
\newblock {\em Surface Science Reports}, 39(7--8):181--235, 2000.

\bibitem{Cohen199042}
M.~L. Cohen and W.~D. Knight.
\newblock {The Physics of Metal Clusters}.
\newblock {\em Phys. Today}, 43(12):42--50, 1990.

\bibitem{Heer1993611}
W.~A. de~Heer.
\newblock The physics of simple metal clusters: experimental aspects and simple
  models.
\newblock {\em Rev. Mod. Phys.}, 65(3):611--676, 1993.

\bibitem{Dymnikov2011901}
V.~D. Dymnikov.
\newblock Fermi energy of electrons in a thin metallic plate.
\newblock {\em Physics of the Solid State}, 53(5):901--907, 2011.

\bibitem{Yong2009155404}
Y.~Han and D.-J. Liu.
\newblock {Quantum size effects in metal nanofilms: Comparison of an
  electron-gas model and density functional theory calculations}.
\newblock {\em Phys. Rev. B}, 80(15):155404, 2009.

\bibitem{Huntington19511035}
H.~B. Huntington.
\newblock {Calculations of Surface Energy for a Free-Electron Metal}.
\newblock {\em Phys. Rev.}, 81(6):1035--1039, 1951.

\bibitem{Kostrobij201651}
P.~Kostrobij and B.~Markovych.
\newblock Effective inter-electron interaction for metallic slab.
\newblock {\em Mathematical Modeling and Computing}, 3(1):51--58, 2016.

\bibitem{Kostrobij2015075441}
P.~P. Kostrobij and B.~M. Markovych.
\newblock {Semi-infinite jellium: Thermodynamic potential, chemical potential,
  and surface energy}.
\newblock {\em Phys. Rev. B}, 92(7):075441, 2015.

\bibitem{Kostrobij2016155401}
P.~P. Kostrobij and B.~M. Markovych.
\newblock {Semi-infinite jellium: Step potential model}.
\newblock {\em Phys. Rev. B}, 93(15):155401, 2016.

\bibitem{Milun200299}
M.~Milun, P.~Pervan, and D.~P. Woodruff.
\newblock Quantum well structures in thin metal films: simple model physics in
  reality?
\newblock {\em Reports on Progress in Physics}, 65(2):99--141, 2002.

\bibitem{Paskin1965A1965}
A.~Paskin and A.~D. Singh.
\newblock {Boundary Conditions and Quantum Effects in Thin Superconducting
  Films}.
\newblock {\em Phys. Rev.}, 140(6A):A1965--A1967, 1965.

\bibitem{Pitarke2001045116}
J.~M. Pitarke and A.~G. Eguiluz.
\newblock Jellium surface energy beyond the local-density approximation:
  Self-consistent-field calculations.
\newblock {\em Phys. Rev. B}, 63(4):045116, 2001.

\bibitem{Schulte1976427}
F.~Schulte.
\newblock A theory of thin metal films: electron density, potentials and work
  function.
\newblock {\em Surface Science}, 55(2):427 -- 444, 1976.

\bibitem{Schulte1977149}
F.~K. Schulte.
\newblock Energies and fermi level of electrons in thin size-quantized metal
  films.
\newblock {\em Physica Status Solidi (b)}, 79(1):149--153, 1977.

\bibitem{Smith1965210}
B.~Smith.
\newblock A size effect in nearly free electron metals.
\newblock {\em Physics Letters}, 18(3):210 -- 211, 1965.

\bibitem{Stratton1965556}
R.~Stratton.
\newblock Disproof of a predicted ``\ldots size effect in nearly free electron
  metals''.
\newblock {\em Physics Letters}, 19(7):556--558, 1965.

\bibitem{Thompson19636}
C.~Thompson and J.~Blatt.
\newblock {Shape resonances in superconductors -- II simplified theory}.
\newblock {\em Physics Letters}, 5(1):6--9, 1963.

\bibitem{Tringides200750}
M.~C. Tringides, M.~Ja{\l}ochowski, and E.~Bauer.
\newblock {Quantum Size Effects in Metallic Nanostructures}.
\newblock {\em Phys. Today}, 60(4):50--54, 2007.

\bibitem{Himbergen19782674}
J.~E. van Himbergen and R.~Silbey.
\newblock Exact solution of metal surface properties in square barrier and
  linear one-electron potential models.
\newblock {\em Phys. Rev. B}, 18(6):2674--2682, 1978.

\bibitem{Biao2008035410}
B.~Wu and Z.~Zhang.
\newblock Stability of metallic thin films studied with a free electron model.
\newblock {\em Phys. Rev. B}, 77(3):035410, 2008.

\end{thebibliography}

\end{document}